\documentclass{jnmp}

\usepackage{amsmath}
\usepackage{epsfig}
\JNMPnumberwithin{equation}{section}

\newtheorem{lemma}{Lemma}[section]
\newtheorem{proposition}{Proposition}[section]

\theoremstyle{definition}

\begin{document}

%  Headings
%
\renewcommand{\evenhead}{A Maccari}
\renewcommand{\oddhead}{The Matrix Kadomtsev--Petviashvili Equation}

%  Titlepage
%
\thispagestyle{empty}

\FirstPageHead{9}{1}{2002}{\pageref{maccari-firstpage}--\pageref{maccari-lastpage}}{Letter}
%  Parameters: Volume, number, year, page range, paper type
%  'Article' could be changed to 'Letter' or 'Review Article'

\copyrightnote{2002}{A Maccari}

\Name{The Matrix Kadomtsev--Petviashvili Equation\\
as a Source of Integrable Nonlinear Equations}
\label{maccari-firstpage}

\Author{Attilio MACCARI}

\Address{Technical Institute ``G. Cardano'', Piazza della Resistenza 1,\\
00015 Monterotondo, Rome, Italy}

\Date{Received May 28, 2001; Revised August 01, 2001; Accepted August 02, 2001}

\begin{abstract}
\noindent
A new integrable class of Davey--Stewartson type systems
of nonlinear partial diffe\-rential equations (NPDEs) in $2+1$ dimensions is
derived from the matrix Kadomtsev--Petviashvili equation by means of an
asymptotically exact nonlinear reduction method based on Fourier expansion
and spatio-temporal rescaling. The integrability by the inverse scattering
method is explicitly demonstrated, by applying the reduction technique also
to the Lax pair of the starting matrix equation and thereby obtaining the
Lax pair for the new class of systems of equations. The characteristics of
the reduction method suggest that the new systems are likely to be of
applicative relevance. A~reduction to a system of two interacting complex
fields is briefly described.
\end{abstract}

\section{Introduction}

New classes of evolution nonlinear partial differential equations (NPDEs)
integrable by the inverse scattering method (S-integrable) have been found
in the last years. These equations are known to be applicable to various
branches of physics such as fluid dynamics, nonlinear optics, condensed
matter physics and so on. The most famous examples are the Korteweg-de Vries
and the nonlinear Schrodinger equations in $1+1$ dimensions and the
Kadomtsev--Petviashvili and the Davey--Stewartson equations in $2+1$ dimensions~[1].

A simple explanation of this coincidence (integrability and applicative
relevance) is based on the observation that very large classes of evolution
NPDEs in $1+1$ and $2+1$ dimensions, with a dispersive linear part, can be
reduced, by a limiting procedure involving the wave modulation induced by
weak nonlinear effects, to a very limited number of ``universal'' evolution
NPDEs. Moreover, the same model equations obtained in this way appear in
many applicative situations (for instance in plasma physics, nonlinear
optics, hydrodynamics, etc.), where weakly nonlinear effects are important
[2--5].

The reduction method preserves integrability and therefore the model
equations are likely to be integrable. For example, it is sufficient that
the very large class of equations from which they are obtainable contains
just one S-integrable equation, provided the limiting procedure preserves
integrability, so that the property of S-integrability is inherited through
this limiting technique. Obviously, the last statement about the
integrability is based on heuristic considerations and could not be
characterized as a rigorous theorem. No precise definition of integrability
is available for evolution NPDEs, there being much difference between
finding the general solution of a NPDE or solving an initial-value problem
with given input data and boundary conditions. It would be possible to
derive the spectral transform of the Davey--Stewartson equation from the
spectral transform of the Kadomtsev--Petviashvili equation.

Thus this approach, besides explaining why certain model equations are
integrable and applicable, provides a powerful tool to investigate the
relation among different integrable equations, to test the integrability of
nonlinear evolution PDEs and, most importantly, to identify integrable
evolution equations that are likely to be of applicative relevance.

In previous papers, we applied this method to certain integrable equations
in $2+1$ dimensions. The most interesting results are that the
Davey-Stewarston equation [6--7] is the typical model equation in $2+1$
dimensions, while new integrable NPDEs can be obtained together with their
Lax pair [8--11]. Moreover, we used the reduction method to derive two
equations of applicative relevance in plasma physics [12--13].

The basic idea of the reduction method is to consider a nonlinear evolution
PDE whose linear part is dispersive; as it is well known the linear
evolution is most appropriately described in terms of Fourier modes and each
Fourier mode evolves with constant amplitude and an associated group
velocity, that represents the speed with which a wave packet peaked at that
Fourier mode would move in configuration space. To evaluate the weak
nonlinear effects it is convenient to consider a specific Fourier mode and
follow it by going over to a frame of reference that moves with its group
velocity. The weak nonlinear effects give rise to a modulation of the
amplitude of that Fourier mode (that would remain constant in the absence of
nonlinear effects). The modulation is best described in terms of rescaled
``coarse-grained'' and ``slow'' variables, that display the weak nonlinear
effects on larger space and time scales; indeed, the first step of the
reduction method is to use a moving frame of reference with the introduction
of the slow variables:
\begin{gather}
\xi = \varepsilon ^{p}(x - V_{1} t),\qquad \eta = \varepsilon ^{p}(y - V_{2} t),
\qquad \tau = \varepsilon ^{q}t, \nonumber\\
 p > 0,\qquad q > 0, \label{maccari:eq1.1}
\end{gather}
where $V_{1} = V_{1} (K_{1} ,K_{2} )$, $V_{2} = V_{2} (K_{1} ,K_{2} )$ are the
components of the group velocity $\underline {V} \left( {\underline {K}}
\right) \equiv \left( V_{1} (K_{1} ,K_{2} ),V_{2} (K_{1} ,K_{2} ) \right)$
of the linearized equation, i.e. of the equation obtained by neglecting all
the nonlinear terms, and $\varepsilon $ is a ``small'' expansion parameter.

It is thereby seen that the function that represents the amplitude
modulation satisfies, in terms of the rescaled, slow, variables, evolution
equations having a universal character; since the coarse-grained nature of
the new variables implies that only certain general features of the
nonlinear interaction are important.

In this paper we expose an interesting extension of this approach and
consider the matrix Kadomtsev--Petviashvili equation [14--15]
\begin{gather}
U_{t} + U_{xxx} - W_{y} + i\sqrt {3} \left[ {W,U} \right] - 3\left\{
{U,U_{x}}  \right\} = 0,\nonumber\\
W_{x} = U_{y},\label{maccari:eq1.2}
\end{gather}
where $[A,B]=AB-BA$, $U = U(x,y,t)$, $W = W(x,y,t)$
 are $N  \otimes N$ complex matrices and
the subscripts denote partial differentiation.

By applying the reduction method, a new class of integrable matrix systems
of evolution NPDEs depending on a real parameter $\lambda$ is obtained
\begin{gather}
i\Psi _{\tau}  + L\Psi - \lambda \left[ \Psi ,\Lambda  \right] - \left[
\Omega ,\Psi  \right] + \sqrt {3} \left\{ \Lambda ,\Psi  \right\} -
\left\{ \Phi ,\Psi ^{2} \right\} = 0,\nonumber\\
i\Phi _{\tau}  - L\Phi - \lambda \left[ \Phi ,\Lambda  \right] - \left[
\Omega ,\Phi  \right] - \sqrt {3} \left\{ \Lambda ,\Phi  \right\} +
\left\{ \Psi ,\Phi ^{2} \right\} = 0,\nonumber\\
\left(3 - \lambda ^{2}\right)\Lambda _{\xi}  + 2\lambda \Lambda _{\eta}  - \Omega
_{\eta}  - \sqrt {3} \left\{ \Psi ,\Phi  \right\}_{\xi}  + \left[ \Psi
,\Phi  \right]_{\eta}  + \lambda \left[ \Phi ,\Psi  \right]_{\xi}  = 0,\nonumber\\
\Lambda _{\eta}  = \Omega _{\xi},\label{maccari:eq1.3}
\end{gather}
where $\{ A,B\} = AB + BA$, the linear differential operator
$L$ is given by
\begin{equation}\label{maccari:eq1.4}
L = - \left(3 + \lambda ^{2}\right){\frac{{\partial ^{2}}}{{\partial \xi ^{2}}}} +
2\lambda {\frac{{\partial ^{2}}}{{\partial \xi \partial \eta} }} -
{\frac{{\partial ^{2}}}{{\partial \eta ^{2}}}},
\end{equation}
and $\Psi = \Psi (\xi ,\eta ,\tau )$, $\Phi = \Phi (\xi ,\eta ,\tau )$,
$\Lambda = \Lambda (\xi ,\eta ,\tau )$ and $\Omega = \Omega (\xi ,\eta ,\tau
)$are $N  \otimes N$ complex matrices.

The paper is organized as follows. In the next section we apply the
reduction method to the starting equation (\ref{maccari:eq1.2}) and obtain the new system of
matrix equations (\ref{maccari:eq1.3})--(\ref{maccari:eq1.4}). Moreover, we reduce the matrix system of
equations to a new integrable two-component complex fields system of
nonlinear equations, which, in the one-component case, reduces to the
standard Davey--Stewartson equation. In Section~3 we discuss in some detail
how the reduction method can be applied to the Lax pair of the equation
(\ref{maccari:eq1.2}) and we derive the Lax pair of the system of matrix equations
(\ref{maccari:eq1.3})--(\ref{maccari:eq1.4}). Finally in the last section we recapitulate the most important
results and indicate some possible extensions.

\section{A new integrable matrix system in \mbox{\mathversion{bold}$2+1$} dimensions}

\begin{lemma}
The linear dispersive part of the starting equation (\ref{maccari:eq1.2}) admits
as a solution a Fourier mode, with a group velocity
 $\underline {V} (\underline {K} ) = \left( V_{1} (K_{1} ,K_{2} ), V_{2}
(K_{1} ,K_{2} ) \right)$,
\begin{gather*}
V_{1} (K_{1} ,K_{2} ) = - 3K_{1}^{2} + \frac{K_{2}^{2}}{K_{1}^{2}},%\label{maccari:eq2.1}\\
\qquad V_{2} (K_{1} ,K_{2} ) = - 2{\frac{{K_{2}} }{{K_{1}} }},\label{maccari:eq2.2}
\end{gather*}
where
\begin{equation*}*\label{maccari:eq2.3}
\underline {V} (\underline {K} ) = {\frac{{\partial \omega} }{{\partial
\underline {K}} }}
\end{equation*}
and $\omega = \omega (K_{1} ,K_{2} ) = - K_{1}^{3} - {\frac{{K_{2}^{2}} }{{K_{1}
}}}$ is the dispersion relation.
\end{lemma}

\begin{proof}
It is sufficient to substitute the plane wave into the linear part of the
matrix KP equation.
\end{proof}

We use the transformation (\ref{maccari:eq1.1}) and introduce the following formal
asymptotic Fourier expansion
\begin{equation}\label{maccari:eq2.4}
U(x,y,t) = {\sum\limits_{n = - \infty} ^{ + \infty}  {\varepsilon ^{\gamma
_{n}} \psi _{n} (\xi ,\eta ,\tau ;\varepsilon )\exp {\left\{ {i(nz)}
\right\}}}},
\end{equation}
where $z = K_{1} x + K_{2} y - \omega t$, $\gamma _{n} = {\left| {n}
\right|}$ for $n \ne 0$, and $\gamma _{0} = r$ is a non negative rational
number which will be fixed later. The unknown functions $\psi _{n} $ depend
on $\varepsilon $ and it is supposed that their limit for $\varepsilon
\to 0$ exists and is finite; in the following this limit will be denoted
with $\psi _{n} (\xi ,\eta ,\tau )$. Moreover we suppose that they can be
expanded in power series of , i.e.
\begin{equation*}%\label{maccari:eq2.5}
\psi _{n} (\xi ,\eta ,\tau ;\varepsilon ) = {\sum\limits_{i = 0}^{\infty}
{\varepsilon ^{i}}} \psi _{n}^{(i)} (\xi ,\eta ,\tau ),
\qquad
\psi _{n} (\xi ,\eta ,\tau ) = \psi _{n}^{(0)} (\xi ,\eta ,\tau ).
\end{equation*}
We now introduce an analogous Fourier expansion
\begin{equation}\label{maccari:eq2.6}
W(x,y,t) = {\sum\limits_{n = - \infty} ^{ + \infty}  {\varepsilon ^{\tilde
{\gamma} _{n}} \varphi _{n} (\xi ,\eta ,\tau ;\varepsilon )\exp {\left\{
{i(nz)} \right\}}}}
\end{equation}
and obtain
\begin{equation*}%\label{maccari:eq2.7}
\varphi _{n} = (K_{2} )(K_{1} )^{ - 1}\psi _{n} + O\left(\varepsilon ^{p}\right).
\end{equation*}
In the following for simplicity we use the abbreviations $\psi _{1}^{(0)} =
\Psi $, $\psi _{ - 1}^{(0)} = \Phi $, $\psi _{0}^{(0)} = \Lambda $ (and
$\phi _{n}^{(0)} = \phi _{n} $, $\phi _{0}^{(0)} = \Omega $).

The final goal is to obtain the evolution equation satisfied by the
modulation amplitudes $\Psi = \Psi (\xi ,\eta ,\tau )$ and $\Phi (\xi ,\eta ,\tau )$
and to understand how it is modified by choosing different wave
numbers.

\begin{proposition}
The matrix system (\ref{maccari:eq1.3})--(\ref{maccari:eq1.4})
can be obtained applying the reduction technique to the matrix
 Kadomtsev-Petviashvili system (\ref{maccari:eq1.2}).
\end{proposition}

\begin{proof}
We insert the expansions (\ref{maccari:eq2.4}) and (\ref{maccari:eq2.6}) into the equation (\ref{maccari:eq1.2}) and
consider the different equations obtained by considering the coefficients of
the Fourier modes.

{\samepage
It is convenient to separate the contributions of the linear and nonlinear
parts by writing
\begin{equation*}%\label{maccari:eq2.8}
\varepsilon ^{\gamma _{n}} D_{n} \psi _{n} = \varepsilon ^{2}F_{n},
\end{equation*}
where $D_{n} $ is a linear differential operator acting on $\psi _{n} (\xi ,\eta ,\tau )$
and $F_{n} $ is the contribution of the nonlinear part. The
operator $D_{n}$ is
\begin{gather*}
D_{n} = \left( - in\omega + \varepsilon ^{q}\partial _{\tau}  - V_{1}
\varepsilon ^{p}\partial _{\xi}  - V_{2} \varepsilon ^{p}\partial _{\eta} \right)
+ \left(inK_{1} + \varepsilon ^{p}\partial _{\xi}  \right)^{3}
- \left(inK_{2} + \varepsilon ^{p}\partial _{\eta}  \right)\nonumber\\
\qquad {} - (i / K_{1} )\left(\varepsilon
^{p}\partial _{\eta}  - (K_{2} / K_{1} )\partial _{\xi}  \right)
 + \left(1 / K_{1}^{2} \right)\varepsilon ^{2p}
\left(\partial _{\xi \eta}  - (K_{2} / K_{1} )\partial _{\xi \xi}  \right).%\label{maccari:eq2.9}
 \end{gather*}
$F_{n} $ can be derived, by assessing the importance of the different terms,
which originate from the nonlinear interaction of the Fourier amplitudes
$\psi _{n} (\xi ,\eta ,\tau )$:
\begin{gather*}%\label{maccari:eq2.10}
F_{2} = 6iK_{1} \Psi ^{2} + O\left(\varepsilon ^{p}\right),\\
F_{0} = \varepsilon ^{p}\left(3{\left\{ {\Psi ,\Phi}  \right\}}_{\xi}  - i\sqrt
{3} {\left[ {\varphi _{1}^{(p)} ,\Phi}  \right]} - i\sqrt {3} {\left[
{\varphi _{ - 1}^{(p)} ,\Psi}  \right]}\right) + O\left(\varepsilon ^{2p},\varepsilon
^{2}\right),\\
F_{1} = \varepsilon ^{r - 1}\left( - i\sqrt {3} {\left[ {\Omega ,\Psi}  \right]}
- i\sqrt {3} {\frac{{K_{2}} }{{K_{1}} }}{\left[ {\Psi ,\Lambda}  \right]} +
3iK_{1} {\left\{ {\Lambda ,\Psi}  \right\}}\right) \nonumber\\
\qquad {}+ 3i\varepsilon K_{1} {\left\{ {\psi _{2} ,\Phi}  \right\}}
+ O\left(\varepsilon ^{r + p - 1},\varepsilon ^{3}\right),
\end{gather*}
and so on.}

By setting $q = 2$, $p = 1$, $r = 2$ for the proper balance of terms, we obtain
the equations for the Fourier modes at the lowest order for $n = 0$, $n = 1$ and $n = 2$:
\begin{gather*}%\label{maccari:eq2.11}
\psi _{2} = - {\frac{{1}}{{K_{1}^{2}} }}\Psi ^{2},\\
\left(3K_{1}^{2} - {\frac{{K_{2}^{2}} }{{K_{1}^{2}} }}\right)\Lambda _{\xi}  +
2{\frac{{K_{2}} }{{K_{1}} }}\Lambda _{\eta}  - \Omega _{\eta}  - 3{\left\{
{\Psi ,\Phi}  \right\}}_{\xi}  \nonumber\\
\qquad {}+ i\sqrt {3} {\left[ {\varphi _{1}^{(p)}
,\Phi}  \right]} + i\sqrt {3} {\left[ {\varphi _{ - 1}^{(p)} ,\Psi}  \right]} = 0,\\
 \Psi _{\tau}  + i\left(3K_{1} + {\frac{{K_{2}^{2}} }{{K_{1}^{3}} }}\right)\Psi _{\xi
\xi}  + {\frac{{i}}{{K_{1}} }}\Psi _{\eta \eta}  - 2i{\frac{{K_{2}
}}{{K_{1}^{2}} }}\Psi _{\xi \eta}  \nonumber\\
\qquad {}+ i\sqrt {3} {\left[ {\Omega ,\Psi}  \right]} + i\sqrt {3} {\frac{{K_{2}
}}{{K_{1}} }}{\left[ {\Psi ,\Lambda}  \right]} - 3iK_{1} {\left\{ {\Lambda
,\Psi}  \right\}} - 3iK_{1} {\left\{ {\psi _{2} ,\Phi}  \right\}} = 0.
 \end{gather*}

Finally, after the cosmetic rescaling
\begin{gather}
\sqrt {3} K_{1} \Lambda \to \Lambda, \qquad
\sqrt {3} \Omega \to \Omega, \qquad
\sqrt {{\frac{{3}}{{K_{1}} }}} \Psi \to \Psi ,\nonumber\\
\sqrt {{\frac{{3}}{{K_{1}} }}} \Phi \to \Phi ,
\qquad
\lambda = {\frac{{K_{2}} }{{K_{1}^{2}} }},
\qquad
{\xi} ' = \xi / \sqrt {K_{1}} , \qquad
{\eta} ' = K_{1} \eta,\label{maccari:eq2.12}
\end{gather}
we arrive at the matrix system of nonlinear evolution equations (\ref{maccari:eq1.3})--(\ref{maccari:eq1.4}).
\end{proof}

This matrix system must be integrable by the spectral transform, because it
has been derived from an S-integrable equation. This is explicitly
demonstrated in the next section.

Let us now look in more detail at the integrable NPDEs implied these
results. If we take $\Phi = \Psi ^{ *} $, $N=1$, we obtain the equation
\begin{equation}\label{maccari:eq2.13}
i\Psi _{\tau}  + L_{1} \Psi + \chi \Psi = 0,
\qquad
L_{2} \chi = 2L_{1} {\left| {\Psi}  \right|}^{2},
\end{equation}
with
\begin{gather*}
\chi _{\eta}  = 2{\left| {\Psi}  \right|}_{\eta} ^{2} + 2\sqrt {3} \Omega _{\xi},\\
L_{1} = - \left(3 + \lambda ^{2}\right){\frac{{\partial ^{2}}}{{\partial \xi ^{2}}}} +
2\lambda {\frac{{\partial ^{2}}}{{\partial \xi \partial \eta} }} -
{\frac{{\partial ^{2}}}{{\partial \eta ^{2}}}},\\
L_{2} = \left(\lambda ^{2} - 3\right){\frac{{\partial ^{2}}}{{\partial \xi ^{2}}}} -
2\lambda {\frac{{\partial ^{2}}}{{\partial \xi \partial \eta} }} -
{\frac{{\partial ^{2}}}{{\partial \eta ^{2}}}}.
\end{gather*}
The NPDE (\ref{maccari:eq2.13}), up to trivial rescalings, coincides with the
Davey--Stewartson equation~[6], whose integrability is well known [7]. Note
that the S-integrable equations found in [10--11] are different from the
standard Davey--Stewartson equation and then not connected with the
S-integrable system (\ref{maccari:eq1.3})--(\ref{maccari:eq1.4}).

In the case $N=2$, we get a nonlinear system for eight interacting fields.
However, an interesting reduction is possible, if we set
\begin{equation*}%\label{maccari:eq2.14}
\Psi = \left(\!\begin{array}{cc}
 {\psi _{1}}  & {\psi _{2}}  \\
 {\psi _{2}}  & {\psi _{1}}
\end{array} \!\right)\!,
\quad
\Phi = \left(\!\begin{array}{cc}
 {\varphi _{1}}  & {\varphi _{2}} \\
 {\varphi _{2}}  & {\varphi _{1}}
\end{array} \!\right)\!,
\quad
\Lambda = \left(\! \begin{array}{cc}
 {\Lambda _{1}}  & {\Lambda _{2}}\\
 {\Lambda _{2}}  & {\Lambda _{1}}
\end{array} \!\right)\!,
\quad
\Omega = \left(\!\begin{array}{cc}
 {\Omega _{1}}  & {\Omega _{2}} \\
 {\Omega _{2}}  & {\Omega _{1}}
\end{array} \!\right)\!.
\end{equation*}

From the matrix system (\ref{maccari:eq1.3})--(\ref{maccari:eq1.4}), we obtain $\Psi = \Phi ^{ *} $ and
\begin{gather}
i\psi _{1,\tau}  + L_{1} \psi _{1} + 2\sqrt {3} (\Lambda _{1} \psi _{1} +
\Lambda _{2} \psi _{2} ) - 2\left({\left| {\psi _{1}}  \right|}^{2}\psi _{1} +
\psi _{1}^{ *}  \psi _{2}^{2} + 2\psi _{1} {\left| {\psi _{2}}
\right|}^{2}\right) = 0,\nonumber\\
i\psi _{2,\tau}  + L_{1} \psi _{2} + 2\sqrt {3} (\Lambda _{1} \psi _{2} +
\Lambda _{2} \psi _{1} ) - 2\left({\left| {\psi _{2}}  \right|}^{2}\psi _{2} +
\psi _{1}^{2} \psi _{2}^{ *}  + 2\psi _{2} {\left| {\psi _{1}}
\right|}^{2}\right) = 0,\nonumber\\
L_{2} \Lambda _{1} = L_{3} \left({\left| {\psi _{1}}  \right|}^{2} + {\left|
{\psi _{2}}  \right|}^{2}\right),
\qquad
L_{2} \Lambda _{2} = L_{3} \left(\psi _{1} \psi _{2}^{ *}  + \psi _{2} \psi _{1}^{ *}  \right),\label{maccari:eq2.15}
\end{gather}
where
\begin{gather*}
L_{1} = - \left(3 + \lambda ^{2}\right)\partial _{\xi} ^{2} + 2\lambda \partial _{\xi
\eta} ^{2} - \partial _{\eta} ^{2} ,\\
L_{2} = \left(3 - \lambda ^{2}\right)\partial _{\xi} ^{2} + 2\lambda \partial _{\xi
\eta} ^{2} - \partial _{\eta} ^{2},
\qquad
L_{3} = 2\sqrt {3} \partial _{\xi} ^{2}.
\end{gather*}

Integrable Davey--Stewartson type equations and system of equations have been
extensively investigated by many authors [16--20]. A detailed list of
Davey-Stewartson systems and equations integrable by the inverse scattering
method has been recently given [21]. The system of equations (\ref{maccari:eq2.15}) does not
appear in these papers. We expect that this new system be integrable by the
inverse scattering method, because it has been obtained from an integrable
equation and the property of integrability is expected to be maintained
through the application of the reduction method. The integrability of the
system of equations (\ref{maccari:eq1.3})--(\ref{maccari:eq1.4}), and of the system (\ref{maccari:eq2.15}) which is a
particular case, is demonstrated in the next section.

\section{The Lax pair for the integrable system of equations}

In this section we apply the reduction method also to the Lax pair of the
starting matrix equation (\ref{maccari:eq1.2}), to demonstrate explicitly the integrability
by the spectral transform of the matrix system (\ref{maccari:eq1.3})--(\ref{maccari:eq1.4}), and we thereby
identify the Lax pair for the system of equations (\ref{maccari:eq1.3})--(\ref{maccari:eq1.4}).

\begin{lemma} The Lax operators $(L, A)$ of the matrix KP equation are
\begin{gather}\label{maccari:eq3.1}
L = {\frac{{i}}{{\sqrt {3}} }}{\frac{{\partial} }{{\partial y}}} +
{\frac{{\partial ^{2}}}{{\partial x^{2}}}} - U(x,y,t),
\qquad
L\phi (x,y,t) = 0,\\
\label{maccari:eq3.2}
A = 4{\frac{{\partial ^{3}}}{{\partial x^{3}}}} - 6U(x,y,t){\frac{{\partial
}}{{\partial x}}} - 3U_{x} (x,y,t) + i\sqrt {3} W,
\end{gather}
with
\begin{equation}\label{maccari:eq3.3}
\phi _{t} (x,y,t) + A\phi (x,y,t) = 0.
\end{equation}
\end{lemma}

\begin{proof}
It can be verified by direct substitution that the operator relation
\begin{equation*}%\label{maccari:eq3.4}
L_{t} = i[L,A] = i(LA - AL)
\end{equation*}
reproduces equation (\ref{maccari:eq1.2}).
\end{proof}

\begin{proposition}
The matrix system (\ref{maccari:eq1.3})--(\ref{maccari:eq1.4}) is S-integrable and its Lax pair
$(L, A)$ is
\begin{equation}\label{maccari:eq3.5}
L\hat {\phi}  = 0,
\end{equation}
where
\begin{gather}
L = \left(\begin{array}{cc}
 {L_{11}}  & {L_{12}}  \\
 {L_{21}}  & {L_{22}}
\end{array}\right),\qquad
L_{11} = I\left(i\partial _{\eta}  + i(\sqrt {3} - \lambda )\partial _{\xi} \right),
\qquad
L_{12} = - \Psi ,\nonumber\\
L_{21} = - \Phi ,
\qquad
L_{22} = I\left(i\partial _{\eta}  - i(\sqrt {3} + \lambda )\partial _{\xi}  \right),\qquad
\hat {\phi}  = \left(\begin{array}{c}
 {\phi _{ +} }  \\
 {\phi _{ -} }
\end{array} \right),\label{maccari:eq3.6}
\end{gather}
$I$ is the $N \otimes N$ unit matrix and
\begin{gather}\label{maccari:eq3.7}
A = \left(\begin{array}{cc}
 {A_{11}}  & {A_{12}}  \\
 {A_{21}}  & {A_{22}}
\end{array} \right),
\end{gather}
where
\begin{gather*}
A_{11} = 6i\partial _{\xi} ^{2} I + i\Omega - i\left(\sqrt {3} + \lambda \right)\Lambda
+ i\left(1 + {\frac{{\lambda \sqrt {3}} }{{12}}}\right)\Phi \Psi ,\\
A_{12} = - 2\sqrt {3} \Psi \partial _{\xi}  - \left(\sqrt {3} + \lambda \right)\Psi
_{\xi}  + \Psi _{\eta}  ,\\
A_{12} = - 2\sqrt {3} \Phi \partial _{\xi}  - \left(\sqrt {3} - \lambda\right )\Phi
_{\xi}  - \Phi _{\eta}  ,\\
A_{22} = - 6iI\partial _{\xi} ^{2} + i\Omega + i\left(\sqrt {3} - \lambda
\right)\Lambda + i\left({\frac{{\lambda \sqrt {3}} }{{12}}} - 1\right)\Psi \Phi .
\end{gather*}
\end{proposition}

\begin{proof}
Let us apply the reduction method to the Lax pair (\ref{maccari:eq3.1})--(\ref{maccari:eq3.3}) of equation
(\ref{maccari:eq1.2}).

The components $\phi _{j} (x,y,t)$, $j=1,\ldots, N$,
of the column vector $\phi (x,y,t)$
can be expanded in Fourier modes as follows
\begin{equation}\label{maccari:eq3.8}
\phi _{j} (x,y,t) = {\sum\limits_{n = - \infty} ^{ + \infty}  {\varepsilon
^{\gamma _{n}} \phi _{j,n} (\xi ,\eta ,\tau ;\varepsilon )\exp {\left[
{i\left( {(\lambda _{1} x + \lambda _{2} y + \lambda _{3} t) +
{\frac{{n}}{{2}}}z} \right)} \right]}}},
\end{equation}
where $z = K_{1} x + K_{2} y - \omega t$, the $\phi _{j,n} (\xi ,\eta ,\tau
;\varepsilon )$ depend parametrically on $\varepsilon $ and remain finite
when $\varepsilon \to 0$, the $\gamma _{n} $ are non negative rational
numbers and $\lambda _{m}$,  $m = 1,\ldots,3$, are real constants to be
determined.

Inserting now the expression for $\phi _{j} (x,y,t)$ in (\ref{maccari:eq3.1}), we derive a
series of relations which are generated by the coefficients of the Fourier
modes. Each relation must be valid for a given order of approximation in
$\varepsilon $.

In particular, for the fundamental harmonics $n = \pm 1,$considering terms
$O\left(\varepsilon ^{0}\right)$ in (\ref{maccari:eq3.1}) and (\ref{maccari:eq3.3}), we obtain
\begin{gather*}%\label{maccari:eq3.9}
{\frac{{i}}{{\sqrt {3}} }}\left(\pm {\frac{{iK_{2}} }{{2}}} + i\lambda _{2} \right) +
\left(\pm {\frac{{iK_{1}} }{{2}}} + i\lambda _{1}\right)^{3} = 0,\\
\left( \mp {\frac{{i\omega} }{{2}}} + i\lambda _{3}\right) + 4\left(\pm {\frac{{iK_{1}
}}{{2}}} + i\lambda _{1} \right)^{3} = 0,
\end{gather*}
and then
\begin{equation*}%\label{maccari:eq3.10}
\lambda _{1} = - {\frac{{K_{2}} }{{2K_{1} \sqrt {3}} }},
\qquad
\lambda _{2} = - {\frac{{\sqrt {3}} }{{4}}}\left({\frac{{K_{2}^{2}} }{{3K_{1}^{2}
}}} + K_{1}^{2} \right),
\qquad
\lambda _{3} = - {\frac{{K_{2}^{3}} }{{6K_{1}^{3} \sqrt {3}} }} -
{\frac{{\sqrt {3}} }{{2}}}K_{1} K_{2}.
\end{equation*}

We thereby understand that the harmonics
\begin{equation}\label{maccari:eq3.11}
\phi _{j,1} ,\qquad \phi _{j, - 1}, \qquad j=1,\ldots,N,
\end{equation}
are fundamental, i.e. for them $\gamma _{n} $ assumes the smallest value,
$\gamma _{n} = 0$.

The successive order $\varepsilon $ for the equation (\ref{maccari:eq3.1}) allow us to obtain the new
spectral problem, because all the $\phi _{j,n} $ may be expressed by means
of the fundamental harmonics (\ref{maccari:eq3.11}), which are connected through the
relations:
\begin{gather}\label{maccari:eq3.12}
i\phi _{ + ,\eta}  + i\left(\sqrt {3} - \lambda \right)\phi _{ + ,\xi}  - \Psi \phi _{ -}  = 0,\\
i\phi _{ - ,\eta}  - i\left(\sqrt {3} + \lambda\right)\phi _{ - ,\xi}  - \Phi \phi _{ +}  = 0,\label{maccari:eq3.13}
\end{gather}
where we set $(\phi _{j,1};\ j = 1,\ldots,N) = \phi _{ +} $,
$(\phi _{j, - 1} ;\ j = 1,\ldots,N) = \phi _{ -}$.

By means of the variable rescaling (\ref{maccari:eq2.12}), and by introducing the $2N  \otimes
2N$ matrix operator $L$, we arrive at the final form (\ref{maccari:eq3.5})--(\ref{maccari:eq3.6}).

To calculate the temporal evolution, we must insert the expression (\ref{maccari:eq3.8}) in
(\ref{maccari:eq3.3}) and consider the relation obtained for the different harmonics $n$ and
for a given order of approximation in $\varepsilon $. If we consider the first order in
$\varepsilon $, we obtain again the spectral problem (\ref{maccari:eq3.5})--(\ref{maccari:eq3.6}).
Only if we take into
account the next orders of approximation of equation (\ref{maccari:eq3.3}), i.e. the order
$\varepsilon ^{2}$, the temporal evolution can be determined. However, new
quantities, the corrections $\tilde {\phi} _{\pm}  (\xi ,\eta ,\tau )$ of
order $\varepsilon $ to the fundamental harmonics $\phi _{\pm}  (\xi ,\eta ,\tau )$,
appear. These unknown quantities can be eliminated in the equation (\ref{maccari:eq3.3}) by
taking advantage of the relation obtained from equation (\ref{maccari:eq3.1}), considering
terms of order $\varepsilon ^{2}$. This elimination is possible only because
equations (\ref{maccari:eq3.1}) and (\ref{maccari:eq3.3}) are identical at the order $\varepsilon $. In
particular, if we consider (\ref{maccari:eq3.3}) calculated to the order $\varepsilon ^{2}$
for $n = \pm 1$, we get
\begin{gather*}\label{maccari:eq3.14}
 \phi _{ + ,\tau}  + 12i\left({\frac{{K_{1}} }{{2}}} + \lambda _{1} \right)\phi _{ +
,\xi \xi}  - 6\Psi \phi _{ - ,\xi}  - 3\Psi _{\xi}  \phi _{ -}
 + {\frac{{\sqrt {3}} }{{K_{1}} }}\left(\Psi _{\eta}  - {\frac{{K_{2}} }{{K_{1}
}}}\Psi _{\xi}  \right)\phi _{ -}  \nonumber\\
\qquad {} + i\sqrt {3} \Omega- 6i\Lambda \left({\frac{{K_{1} }}{{2}}} + \lambda _{1} \right)\phi _{ +}
 + i\left( - 6\lambda _{1} - 6K_{1} + \sqrt {3} {\frac{{K_{2}} }{{K_{1}} }}\right)\Phi
\phi _{ + 3} \nonumber\\
\qquad {} - {\frac{{2iK_{2} \sqrt {3}} }{{K_{1}} }}\left({\frac{{i}}{{\sqrt {3}} }}\tilde
{\phi} _{ + ,\eta}  + i\left(K_{1} - {\frac{{K_{2}} }{{K_{1} \sqrt {3}} }}\right)\tilde
{\phi} _{ + ,\xi}  - \Psi \tilde {\phi} _{ -}  \right) = 0, \\
 \phi _{ - ,\tau}  + 12i\left( - {\frac{{K_{1}} }{{2}}} + \lambda _{1} \right)\phi _{ -
,\xi \xi}  - 6\Phi \phi _{ + ,\xi}  - 3\Phi _{\xi}  \phi _{ +}
 - {\frac{{\sqrt {3}} }{{K_{1}} }}\left(\Phi _{\eta}  - {\frac{{K_{2}} }{{K_{1}
}}}\Phi _{\xi}  \right)\phi _{ +}  \nonumber\\
\qquad {} + i\sqrt {3} \Omega- 6i\Lambda \left( -
{\frac{{K_{1}} }{{2}}} + \lambda _{1} \right)\phi _{ -}
 + i\left( - 6\lambda _{1} + 6K_{1} + \sqrt {3} {\frac{{K_{2}} }{{K_{1}} }}\right)\Psi
\phi _{ - 3} \nonumber\\
\qquad {}- {\frac{{2iK_{2} \sqrt {3}} }{{K_{1}} }}\left({\frac{{i}}{{\sqrt {3}} }}\tilde
{\phi} _{ - ,\eta}  - i\left(K_{1} + {\frac{{K_{2}} }{{K_{1} \sqrt {3}} }}\right)\tilde
{\phi} _{ - ,\xi}  - \Phi \tilde {\phi} _{ +}  \right) = 0.
 \end{gather*}

To evaluate this expression we took advantage of the fact that $\phi _{\pm
3} $ are connected with the fundamental harmonics (these relations are
obtained from (\ref{maccari:eq3.1}) for $n = \pm 3$ at the lower order in $\varepsilon $):
\begin{equation}\label{maccari:eq3.15}
\phi _{ + 3} = \left( {{\frac{{ - 1}}{{2K_{1}^{2}} }}} \right)\Psi \phi _{ + } ,
\qquad
\phi _{ - 3} = \left( {{\frac{{ - 1}}{{2K_{1}^{2}} }}} \right)\Phi \phi _{ - } .
\end{equation}

We now consider the equation (\ref{maccari:eq3.1}) at the order $\varepsilon ^{2}$ for $n =
\pm 1$, which provides the corrections $\tilde {\phi} _{ +}  (\xi ,\eta
,\tau )$, $\tilde {\phi} _{ -}  (\xi ,\eta ,\tau )$. Via the transformation
(\ref{maccari:eq2.12}) and after a lengthy calculation we arrive at the final form (\ref{maccari:eq3.7}) for
the $2N  \otimes 2N$ matrix operator $A$, which satisfies the equation
\begin{equation}\label{maccari:eq3.16}
\hat {\phi} _{\tau}  + A\hat {\phi}  = 0.
\end{equation}
\end{proof}

The determination of the Lax pair (\ref{maccari:eq3.6}) and (\ref{maccari:eq3.7}), which satisfies the
equations (\ref{maccari:eq3.5}) and (\ref{maccari:eq3.16}), demonstrates the S-integrability of the system
(\ref{maccari:eq1.3})--(\ref{maccari:eq1.4}).

\section{Conclusion}

We have derived a new, integrable, and presumably of applicative interest,
nonlinear mat\-rix system of evolution equations of Davey--Stewartson type from
the integrable matrix equation (\ref{maccari:eq1.2}), by means of an extension of the
reduction method based on Fourier expansion and space-time rescalings. It
reduces to the standard Davey--Stewartson equation in the single mode case
and to a new integrable system of two interacting fields in the $N=2$ case.
Moreover, we have applied the reduction method to the Lax pair (\ref{maccari:eq3.1})--(\ref{maccari:eq3.3}) of
the original equation and have demonstrated the integrability property of
the new matrix system of equations, by exhibiting the corresponding Lax pair
(\ref{maccari:eq3.11})--(\ref{maccari:eq3.12}) and (\ref{maccari:eq3.15})--(\ref{maccari:eq3.16}).

We have outlined the approach that permits to obtain such system of
equations and the next steps will be the explicit resolution of the spectral
problem and the possible identification of localized or asymptotically
finite solutions.

It is also convenient to push the approach beyond its ``leading order''
application by considering different rescalings in the two spatial variables
or looking at special cases when some key parameters vanish, in analogy to
the case of the model equations treated in [8--9].

\label{maccari-lastpage}

\end{document}